# THE MECHANISMS OF ELECTRON ACCELERATION DURING MULTIPLE X LINE MAGNETIC RECONNECTION WITH A GUIDE FIELD

Huanyu Wang[1,2], Quanming Lu[1,2], Can Huang[1,2], Shui Wang[1,2]

[1]CAS Key Lab of Geospace Environment, Department of Geophysics and Planetary Science, University of Science and Technology of China, Hefei 230026, China

[2]Collaborative Innovation Center of Astronautical Science and Technology, China

Corresponding Author: Quanming Lu

Email: qmlu@ustc.edu.cn

# Abstract

The interactions between magnetic islands are considered to play an important role in electron acceleration during magnetic reconnection. In this paper, two-dimensional (2-D) particle-in-cell (PIC) simulations are performed to study electron acceleration during multiple X line reconnection with a guide field. The electrons remain almost magnetized, and we can then analyze the contributions of the parallel electric field, Fermi and betatron mechanisms to electron acceleration during the evolution of magnetic reconnection by comparing with a guide-center theory. The results show that with the proceeding of magnetic reconnection, two magnetic islands are formed in the simulation domain. The electrons are accelerated by both the parallel electric field in the vicinity of the X lines and Fermi mechanism due to the contraction of the two magnetic islands. Then the two magnetic islands begin to merge into one, and in such a process electrons can be accelerated by the parallel electric field and betatron mechanisms. During the betatron acceleration, the electrons are locally accelerated in the regions where the magnetic field is piled up by the high-speed flow from the X line. At last, when the coalescence of the two islands into a big one finishes, electrons can further be accelerated by the Fermi mechanism because of the contraction of the big island. With the increase of the guide field, the contributions of Fermi and betatron mechanisms to electron acceleration become less and less important. When the guide field is sufficiently large, the contributions of Fermi and betatron mechanisms are almost negligible.

# 1. INTRODUCTION

Magnetic reconnection is a fundamental physical process in plasma which is closely related to rapid energy conversion. In magnetic reconnection, free magnetic energy stored in a current sheet is suddenly released, and the plasma is then accelerated and heated (Vasyliunas 1975; Biskamp 2000; Priest & Forbes 2000; Birn et al. 2001; Daughton et al. 2006; Lu et al. 2013). Accelerated electrons during magnetic reconnection are thought to provide the non-thermal part of electron spectra observed in many explosive phenomena such as solar flares (Lin et al. 1976, 2003; Miller et al. 1997), substorms in the Earth`s magnetosphere (Øieroset et al. 2002; Imada et al. 2007; Wang et al. 2010), and disruption in laboratory fusion experiments (Wesson 1997; Savrukhin 2001). For example, x rays observed in solar flares are thought to be generated by the energetic electrons accelerated during magnetic reconnection (Yokoyama & Shibata 1995; Manoharan et al. 1996; Longcope et al, 2001). However, how energetic electrons are produced during magnetic reconnection is a long-standing problem, which is getting more and more attention recently. Many theoretical efforts have been devoted to reveal the mechanisms of electron acceleration during magnetic reconnection (Hoshino et al., 2001; Fu et al., 2006; Drake et al., 2006; Huang et al., 2010, 2015; Guo et al., 2014).

Electron acceleration by the reconnection electric field in the vicinity of the X line was previously thought the primary mechanism during magnetic reconnection. In anti-parallel magnetic reconnection, electrons meander through the vicinity of the X line, and are accelerated by the reconnection electric field (Vasyliunas, 1975; Litvinenko, 1996; Hoshino et al. 2001; Fu et al. 2006; Huang et al. 2010). In guide field reconnection, electrons may be pre-accelerated by the parallel electric field in the separatrix region before they enter the vicinity of the X line (Drake et al 2005; Pritchett 2006; Egedal et al. 2013), where these electrons stay a longer time due to the gyration in the guide field (Fu et al. 2006; Huang et al 2010). In this way, the efficiency of electron acceleration in the vicinity of the X line may be enhanced in guide field reconnection. Hoshino et al. (2001) demonstrated that electrons can be further accelerated stochastically by the reconnection electric field after they enter the pileup region. The jet front driven by an ion outflow is another site to accelerate electrons, where the electrons are highly energized in the perpendicular direction due to the betatron acceleration (Fu et al. 2011; Huang et al. 2012; Birn et al. 2013; Wu et al. 2013). The parallel electric field is considered to play an important role in trapping these energetic electrons in the jet front region, then the electrons are energized due to the betatron acceleration (Huang et al. 2015). Besides, magnetic islands also play a critical role in electron acceleration during magnetic reconnection (Fu et al. 2006; Drake et al. 2006; Pritchett 2008; Chen et al. 2008; Oka et al. 2010; Guo et al. 2014). Fu et al. (2006) and Drake et al. (2006) proposed that electrons can gain energy when they are reflected from the two ends of a contracting magnetic island, which has also been verified by in situ observations (Chen et al. 2008). With in situ observations in the earth's magnetotail, Wu et al. (2015) demonstrated that a multistage is necessary to accelerate electrons to high energy in magnetic reconnection.

With a guiding-center theory, Dahlin et al. (2014) explored the importance of different acceleration mechanisms in guide field reconnection. Under the guiding-center approximation, the evolution of the energy $\varepsilon$ of a single electron can be given as (Northrop 1963; Dahlin et al. 2014)

$$\frac{d\varepsilon}{dt} = (\mu/\gamma)\partial_t B - e(v_\| \boldsymbol{b} + \boldsymbol{v}_c + \boldsymbol{v}_g) \cdot \boldsymbol{E} ,  \qquad (1)$$

where $\boldsymbol{b} = \boldsymbol{B}/|\boldsymbol{B}|$, $\mu = m_e \gamma^2 v_\perp^2 / 2B$ is the magnetic moment, $\gamma$ is the Lorentz factor, and $v_\| = \boldsymbol{v} \cdot \boldsymbol{b}$. $\boldsymbol{v}_c = (v_\|^2 \boldsymbol{b}/\Omega_{ce}) \times \boldsymbol{\kappa}$, and $\boldsymbol{v}_g = (v_\perp^2 \boldsymbol{b}/(2\Omega_{ce})) \times (\nabla B / B)$ are the curvature and gradient $B$ drifts, respectively. $\Omega_{ce} = eB/\gamma m_e c$ is the electron cyclotron frequency, and $\boldsymbol{\kappa} = \boldsymbol{b} \cdot \nabla \boldsymbol{b}$ is the curvature. Eq. (1) can be described as follows after all particles in a local region are summed (Dahlin et al. 2014)

$$\frac{dU}{dt} = E_\parallel J_\parallel + \frac{p_\perp}{B}\left(\frac{\partial B}{\partial t} + \boldsymbol{u}_E \cdot \nabla B\right) + \left(p_\parallel + m_e n u_\parallel^2\right)\boldsymbol{u}_E \cdot \boldsymbol{\kappa}, \qquad (2)$$

where $U$ is the total kinetic energy, $\boldsymbol{u}_E$ is the '$\boldsymbol{E}\times\boldsymbol{B}$' drift velocity, $u_\parallel$ is the bulk velocity parallel to the magnetic field, $n$ is the electron density, $p_\perp$ and $p_\parallel$ are the perpendicular and parallel pressures, respectively. The first term in Eq. (2) is the acceleration by the parallel electric field, and the second term is the betatron mechanism corresponding to perpendicular heating or cooling due to the conservation of magnetic moment $\mu$. The last term drives parallel acceleration, which arises from the first-order Fermi mechanism (Northrop 1961; Drake et al. 2006). Dahlin et al. (2014) found that in magnetic reconnection with a small guide field the Fermi acceleration is the dominant source for electron energization, and with the increase of the guide field electron acceleration by the parallel electric field becomes comparable to that of the Fermi acceleration. Recently, the interactions between magnetic islands (such as merging of islands) have been found to lead to a great enhancement of energetic electrons (Pritchett 2008; Oka et al. 2010; Tanaka et al. 2010; Hoshino et al. 2012; Zank et al. 2014). The electrons are found to be highly accelerated around the merging point of the secondary reconnection during the coalescence of magnetic islands, which is driven by the converging outflows from the initial magnetic reconnection regions (Oka et al., 2010). The current sheet in the solar atmosphere (Sui & Holman 2003; Liu et al. 2010) and the earth's magnetosphere (Deng et al., 2004; Eastwood et al., 2005; Wang et al., 2014) usually have a sufficient length, where occurred magnetic reconnections in general have multiple X lines, and many islands are generated and then interact with each other (Nakamura et al. 2010; Huang et al. 2012; Eriksson et al. 2014). In this paper, with two-dimensional (2D) particle-in-cell (PIC) simulations, electron acceleration during multiple X line reconnection with a guide field is investigated by comparing with a guiding-center theory. We follow the time evolution of electron energy, and their sources in specific flux tubes and over the spatial domain during island generation, during island merging, and after coalescence has completed. The contributions of the parallel electric field, Fermi and betatron mechanisms to electron acceleration at different stages are analyzed in detail, and the effects of the guide field are also studied.

The paper is organized as follows. In Section 2, we delineate our simulation model. The simulation results are presented in Section 3. We summarize our results and discuss their significance in Section 4.

## 2. SIMULATION MODEL

A 2D PIC simulation model is used in this paper to investigate the mechanisms of electron acceleration during the interactions between magnetic islands formed in multiple X line reconnection with a guide field. In our PIC simulations, the electromagnetic fields are defined on the grids and updated by solving the Maxwell equation with a full explicit algorithm, and ions and electrons are advanced in these electromagnetic fields. The initial configuration of the magnetic field consists of a uniform guide field superimposed by a Harris equilibrium. The magnetic field and the corresponding number density are given by

$$\boldsymbol{B}_0(z) = B_0 \tanh(z/\delta)\boldsymbol{e}_x + B_{y0}\boldsymbol{e}_y, \qquad (3)$$

$$n(z) = n_b + n_0 \mathrm{sech}^2(z/\delta), \qquad (4)$$

where $B_0$ is the asymptotic magnetic field, $\delta$ is the half-width of the current sheet, $B_{y0}$ is the initial guide field perpendicular to the reconnection plane, $n_b$ is the number density of the background plasma, and $n_0$ is the peak Harris number density. The initial distribution functions for ions and electrons are Maxwellian with a drift speed in the $y$ direction, and the drift speeds satisfy the following equation: $V_{i0}/V_{e0} = -T_{i0}/T_{e0}$, where $V_{e0}(V_{i0})$ and $T_{e0}(T_{i0})$ are the initial drift speed and the temperature of electrons(ions), respectively. We set $T_{i0}/T_{e0} = 4$, and $n_b = 0.2n_0$ in our simulations. The initial half-width of the current sheet is set to be $\delta = 0.5 d_i$ (where $d_i = c/\omega_{pi}$ is the ion inertial length defined on $n_0$) and the mass ratio $m_i/m_e = 100$. The light speed $c = 15 v_A$, where $v_A$ is the Alfven speed based on $B_0$ and $n_0$.

The computations are carried out in a rectangular domain in the $(x,z)$ plane with the dimension $L_x \times L_z = (51.2 d_i) \times (12.8 d_i)$. The grid number is $N_x \times N_z = 1024 \times 256$. Therefore, the spatial resolution is $\Delta x = \Delta z = 0.05 d_i$. The time step is $\Omega_i t = 0.001$, where $\Omega_i = eB_0/m_i$ is the ion gyro-frequency. We employ more than $10^7$ particles per species to simulate the plasma. Periodic boundary condition for the electromagnetic field and particles along the $x$ axis, and the ideal conducting boundary condition for the electromagnetic field and reflected boundary condition for particles in the $z$ direction are used. The reconnection is initiated by a small flux perturbation same as done in the GEM challenge (Birn et al, 2001) simulations because of the limited computing power. So the reconnection initiates with states similar to that of a spontaneous reconnection except that it bypasses the linear growth rate of the tearing mode.

In order to investigate the mechanisms to produce the non-thermal electrons during the evolution of multiple X line reconnection with a guiding-center theory, we limit our simulations to guide field reconnection. In this paper, we run three cases with the initial guide field $B_{y0} = 0.5 B_0$, $1.0 B_0$ and $2.0 B_0$.

## 3. SIMULATION RESULTS

In order to analyze the mechanisms to electron acceleration, we trace the distributions of electron energy in a defined flux tube during the evolution of multiple X line reconnection, and then calculate the contributions of the parallel electric field, Fermi and betatron mechanisms to electron acceleration, which are based on Eq. (2). Figure 1(a) shows the magnetic field lines and the distributions of electron energy in the defined flux tube at $\Omega_i t = 20, 25, 30, 35, 40$ and $45$, while Figure 1(b) exhibits the evolution of the contributions of the parallel electric field, Fermi and betatron mechanisms to the enhancement of electron energy in the flux tube. Here the initial guide field is $B_{y0} = 0.5B_0$. The reconnection of magnetic field lines begins at about $\Omega_i t = 15$, and the X line appears around the boundary of the simulation domain. At this time, there is one magnetic island in the simulation domain, and the energy of the electrons in the flux tube is enhanced. These electrons are accelerated by the parallel electric field in the vicinity of the X line around the boundary. Simultaneously, each of the flux tubes is contracted due to the compression by the high-speed outflow from the X line, the electrons are also accelerated by the Fermi mechanism. At about $\Omega_i t = 23$, another X line is formed around the center of the simulation domain, and two magnetic islands are formed. The flux tube is separated into two detached tubes, which are contracted due to the compression of the high-speed outflow from the two X lines. The electron energy is further enhanced due to acceleration by both the Fermi mechanism and the parallel electric field in the vicinity of the X line around the center of the simulation domain. Simultaneously, the electrons suffer the betatron cooling because of the annihilation of the magnetic field during magnetic reconnection. Then the electrons are accelerated due to the betatron acceleration when the magnetic field begins to be piled up at the ends of magnetic islands by the high-speed flow from the X lines. From about $\Omega_i t = 33$, the two islands in the simulation domain begins to merge into a big island, and the electrons in the flux tube are accelerated by the parallel electric field around the merging point. After the coalescence is finished, a big island is formed and the flux tubes are merged into a big one. The electrons can also be accelerated by the Fermi mechanism due to the contraction of the flux tube. Note that the betatron acceleration or cooling is a local process, which can only affect the electrons in a region where the magnetic field is piled up or annihilated. Their contributions to the energetic electrons in the whole flux tube is smaller than that of the parallel electric field and Fermi mechanisms.

Figure 2(a) plots the configuration of five different flux tubes with magnetic flux: $\psi \in [0.5B_0 d_i, 1.0B_0 d_i]$, $[1.0B_0 d_i, 1.5B_0 d_i]$, $[1.5B_0 d_i, 2.0B_0 d_i]$, $[2.0B_0 d_i, 2.5B_0 d_i]$ and $[2.5B_0 d_i, 3.0B_0 d_i]$, marked with different colors: blue, green, yellow, red, purple at $\Omega_i t = 20, 25, 30, 35, 40$ and $45$, while Figure 2(b), 2(c) and 2(d) exhibits the evolution of the contributions of the parallel electric field, Fermi and betatron mechanisms to the enhancement of electron energy in different flux tubes, and the sum of these contributions is shown in Figure 2(e). The different colored lines correspond to different flux tubes in Figure 2(a). Similar to Figure 1, the contributions of the betatron acceleration in a whole flux tube is smaller than that of the parallel electric field and Fermi mechanisms due to its local effects. Both the parallel electric field and Fermi mechanism are important to each flux tube, and acceleration efficiency by both the parallel electric field and Fermi mechanism becomes lower when the flux tubes locate far enough away from the center of the current sheet.

In Figure 3, with the same method in Dahlin et al. (2014), from the top to the bottom panel, we plot the spatial distributions of the electron nongyrotropy, the contributions of the parallel electric field, Fermi and betatron mechanisms to the electron acceleration, and the spatially integrated contribution $\Xi(x) = \int_0^x dx' \int U(x', z) dz$ at $\Omega_i t =$ (a)25, (b)35, and (c)45, respectively. Here the initial guide field is $B_{y0} = 0.5B_0$. The electron nongyrotropy is calculated by $D_{ng} = \dfrac{2\sqrt{\sum_{i,j} N_{ij}^2}}{Tr(\mathbf{P}_e)}$, where $\mathbf{P}_e$ is the electron full pressure tensor and $N_{ij}$ are the matrix elements of $\mathbf{N}$, defined as the nongyrotropic part of the electron full pressure tensor (Aunai et al., 2013). In the expression of $\Xi$, $U$ is the term contributed by the parallel electric field, Fermi or the betatron mechanism, based on Eq. (2). Therefore, the slope of $\Xi$ yields the contribution from the corresponding term at a given $x$. The electron nongyrotropy is almost zero in the whole simulation domain except at a small region along the separatrices, and it means that the guide-center theory can be used to analyze the electron acceleration. The time at $\Omega_i t = 25$ represents the stage where two magnetic islands are formed and being contracted by the high-speed flow from the X line, the electrons are mainly accelerated by the parallel electric field and Fermi mechanisms. Although the contributions of the betatron acceleration to electron acceleration cannot be negligible, the betatron acceleration is in general accompanied by the betatron cooling because the pileup and annihilation of magnetic field usually occurs simultaneously during the interactions of magnetic islands, and their net effects to electron acceleration may be smaller than that from the parallel electric field or Fermi mechanism. The time at $\Omega_i t = 35$ is the stage where the two islands are merging, when electrons are accelerated mainly by the parallel electric field. At $\Omega_i t = 45$, a big magnetic island is formed and being contracted after the coalescence of the two islands is finished, and electrons are mainly accelerated by the Fermi mechanism. These results are consistent with the conclusions obtained from Figure 1.

From Figure 3, we further shown that when the two magnetic islands are being contracted (at $\Omega_i t = 25$), the electron acceleration by the parallel electric field occurs mainly in the vicinity of the X line, while the electrons at the two ends of a magnetic island are accelerated due to the Fermi or betatron mechanism. At $\Omega_i t = 35$, where the islands are merging, electrons are mainly accelerated around the merging point by the parallel electric field. At $\Omega_i t = 45$, when the coalescence of the two islands is finished and a big island is formed, electrons are mainly accelerated at the two ends of the big island due to the Fermi mechanism.

Figure 4 plots the evolution of the spatial distribution of electrons with energy larger than $0.1m_ec^2$ and the contributions of the parallel electric field, Fermi and betatron mechanisms to the enhancement of electron energy in the whole simulation domain with different initial guide fields (a) $B_{y0}=0.5B_0$, (b) $B_{y0}=1.0B_0$ and (c) $B_{y0}=2.0B_0$, respectively. The process of electron acceleration can be separated into two stages. In the case with $B_{y0}=0.5B_0$, in the first stage (from about $\Omega_i t=15$ to 30), electrons are accelerated mainly by both the parallel electric field and Fermi mechanisms when the two magnetic island are formed and being contracted. In the second stage (from about $\Omega_i t=30$ to 40), electrons are first accelerated by the parallel electric field induced during the coalescence of the two magnetic island, and then the Fermi mechanism begins to work when the newly formed big island start contracting. In both stages, the electron acceleration by the two mechanisms is comparable, and the net effect of the betatron acceleration is smaller.

For the case with the guide field $B_{y0}=1.0B_0$, in the first stage (from about $\Omega_i t=15$ to 40), two magnetic islands are formed and being contracted, and in the second stage (from about $\Omega_i t=40$ to 60), the two magnetic islands are merged into a big one. The process of electron acceleration is similar to that with the guide field $B_{y0}=0.5B_0$, however, now the acceleration by the parallel electric field is more important than that by the Fermi acceleration. In the case with the guide field $B_{y0}=2.0B_0$, the contribution of the Fermi mechanism can be neglected, although the evolution of the magnetic field lines and electron acceleration also have two stages similar to the results with a small guide field. When we increase the guide field, the energetic electrons tend to gather to the edge of the magnetic island, because the Fermi acceleration becomes less and less important, and the parallel electric field can only accelerate the electrons at the edge of magnetic island. Also in these two cases, the net effect of the betatron acceleration is smaller than that of the parallel electric field or Fermi mechanism. In Figure 4, the difference between $dU/dt$ and the 'sum' comes from the non-adiabatic motion of some electrons, which cannot be described by a guiding-center theory. The difference becomes smaller with the increase of the guide field, because with the increase of the guide field the electron motions can be described more precisely with a guiding-center theory.

Figure 5 shows electron momentum spectra in the directions parallel and perpendicular to the magnetic field during magnetic reconnection with a guide field （a） $B_{y0} = 0.5B_0$, (b) $B_{y0} = 1.0B_0$, (c) $B_{y0} = 2.0B_0$, and the spectra are obtained by integrating all electrons in the simulation domain. Initially, the distribution of these electrons satisfy a Maxwellian function in both the parallel and perpendicular directions. A non-thermal tail of both parallel and perpendicular energy is formed during magnetic reconnection with the parallel momentum larger than the perpendicular momentum, similar to the results in Dahlin et al. (2014). As the parallel electric field and Fermi mechanism are main contributions to electron acceleration during the interaction of magnetic islands, they tend to accelerate electrons in the parallel direction. The electron acceleration in the perpendicular direction may come from the betatron acceleration or from the non-adiabatic motions of the high energy electrons or the electrons traveling through the separatrices.

## 4. DISCUSSION AND CONCLUSIONS

In this paper, with a 2D PIC simulation model we studied electron acceleration during multiple X line reconnection with a guide field by following the time develop of electron energy, and their sources in specific flux tubes and over the spatial domain. The evolution of magnetic reconnection and the associated electron acceleration has two distinct stages. In the first stage, two magnetic islands are formed in the simulation domain, and they are contracted by the high-speed flow produced from the X lines. Electrons can be accelerated in the vicinity of the X line by the parallel electric field, as well as at the two ends of each magnetic island by the Fermi mechanism. During this stage, the contributions of the betatron mechanism to electron acceleration may be also important. However, the betatron mechanism is a local process and only affects the electrons in a region where the magnetic field is piled up. At the same time, the pile up and annihilation of the magnetic field usually occurs simultaneously during the interactions of magnetic island, and the betatron acceleration is in general accompanied by the betatron cooling. Their net effects to electron acceleration may be smaller than that of the parallel electric field or Fermi mechanism. In the second stage, the two magnetic islands are merged into a big one. Electrons are firstly accelerated around the merging point by the parallel electric field, and then are accelerated due to the Fermi mechanism because the big island begins to be contracted after the coalescence is finished. We also changed the size of simulation domain, and find that it doesn't change the relative importance of different acceleration mechanisms after doubling the system size along the x axis. When the guide field is small, the contribution of the Fermi mechanism to electron acceleration is comparable to that of the parallel electric field. However, with the increase of the guide field, the formed magnetic islands become more and more difficult to compress, and then the contribution of the Fermi mechanism becomes less and less important. When the guide field is sufficiently large ($B_{y0} \geq 2.0B_0$), the contributions of the Fermi mechanism to electron acceleration is negligible. When the guide field is sufficiently small ($B_{y0} \leq 0.2B_0$), the Fermi acceleration will become more important than that of the parallel electric field, as describe in Dahlin et al (2014). However, in anti-parallel magnetic reconnection, the motions of most electrons will become non-adiabatic, and a guiding-center theory cannot be used. How to analyze the mechanisms of electron acceleration is such a situation is beyond the scope of this paper.

Energetic electrons are one of the most important signatures in magnetic reconnection. In space plasma, such as in solar atmosphere, a current sheet, where magnetic reconnections occur, usually has a large aspect ratio of the length to the width and a finite guide field (Sui & Holman 2003; Liu et al. 2010). Therefore, magnetic reconnection in such a current sheet usually has multiple X lines, and the interactions between magnetic islands are prevalent (Nakamura et al. 2010; Huang et al 2012; Eriksson et al. 2014). Our simulations have shown that the parallel electric field and Fermi mechanisms provide two important ways to produce these energetic electrons during magnetic reconnection. When the guide field is sufficiently strong, the contribution of the Fermi mechanism to electron acceleration is negligible during the contraction of magnetic island.

## Acknowledgements

This research was supported by 973 Program (2013CBA01503, 2012CB825602), the National Science Foundation of China, Grant Nos. 41474125, 41331067, 41174124, 41121003, 41404129, and CAS Key Research Program KZZD-EW-01-4.

Zank, G. P., le Roux, J. A., Webb, G. M., et al. 2014, Astrophys. J., 797, 28

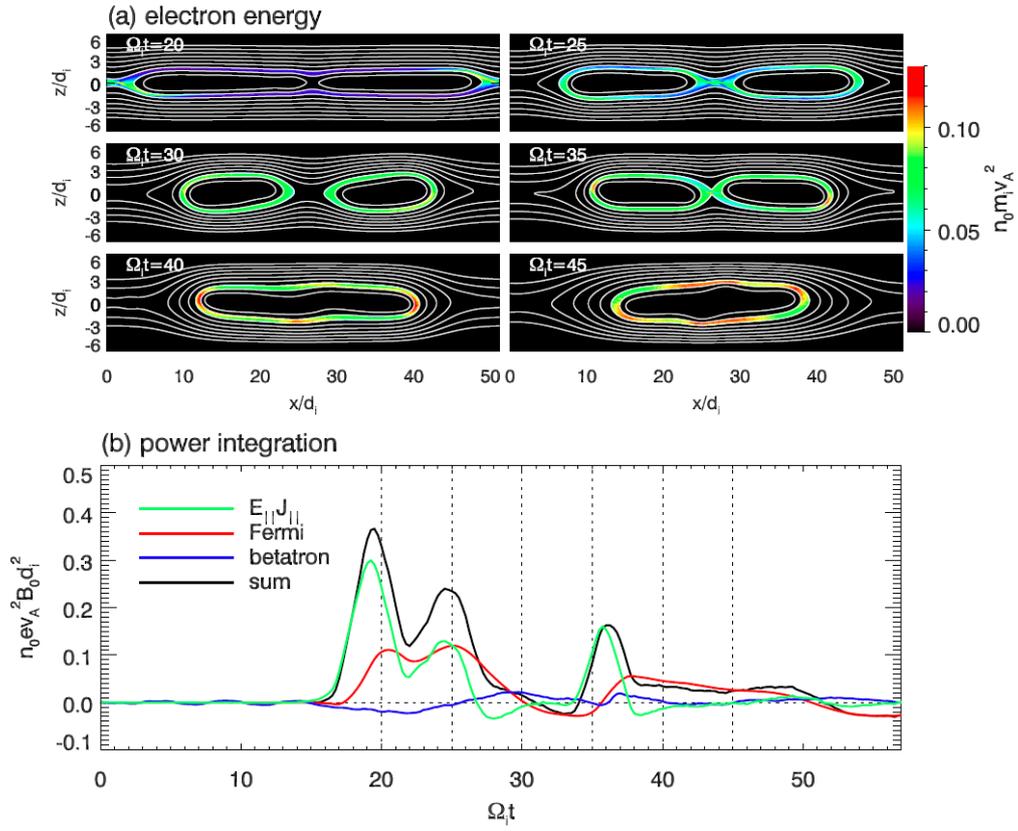

Figure 1. Results from a simulation with a guide field of $0.5B_0$. The time evolution of (a) magnetic field lines and electron energy in the defined flux tube, (b) the contributions of the parallel electric field, Fermi and betatron mechanism to electron acceleration in the flux tube, which is calculated from Eq. (2).

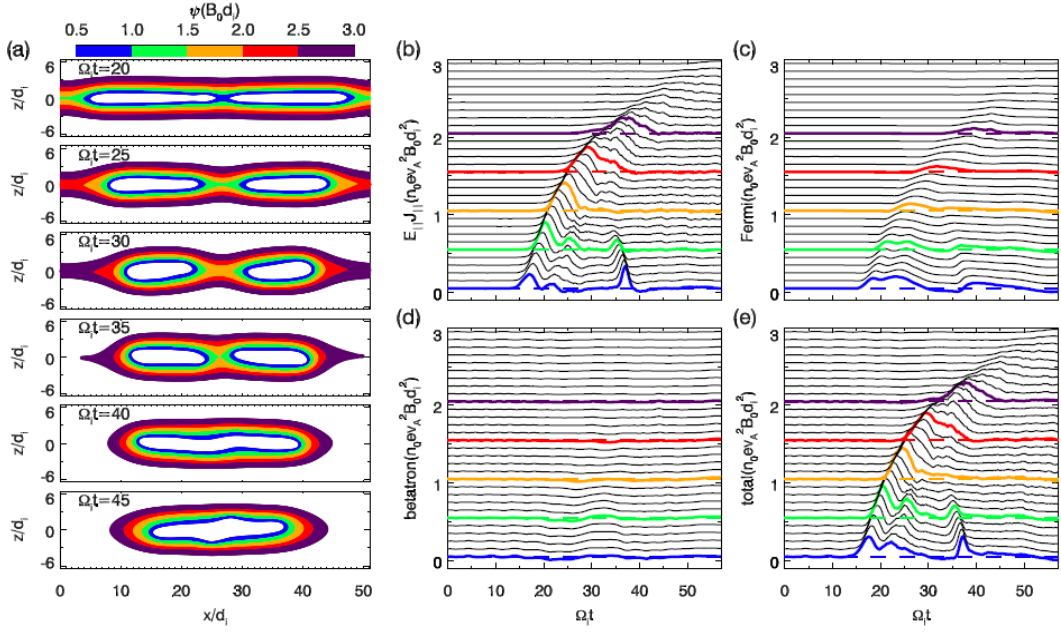

Figure 2. (a)The configuration of different magnetic flux tubes marked in different colors at $\Omega_i t$ = 20, 25, 30, 35, 40 and 45, the flux tube with magnetic flux $\psi \in [0.5B_0 d_i, 1.0B_0 d_i]$ marked in blue, the flux tube with magnetic flux $\psi \in [1.0B_0 d_i, 1.5B_0 d_i]$ marked in green and so on. (b), (c) and (d) are the evolutions of the contributions to the enhancement of electron energy in different flux tubes by the parallel electric field, Fermi and betatron mechanisms, respectively. (e) is the sum of these contributions to electron acceleration. The different colors corresponds to different flux tubes, and the color lines in (b)-(e) denote the flux tubes marked with the same color in (a). The black lines between the colored lines denote the flux tubes among the colored flux tubes shown in (a).

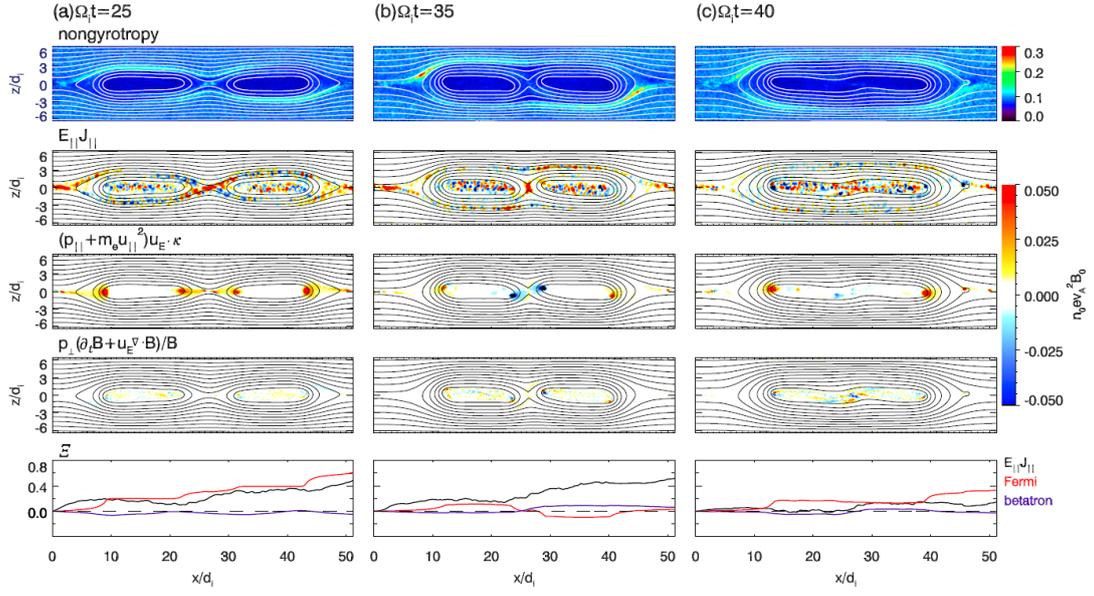

Figure 3. From the top to bottom panels, the spatial distribution of the electrons nongyrotropy, the contributions of the parallel electric field, Fermi and betatron mechanisms, and spatially integrated contribution $\Xi$ for the guide field $0.5B_0$ at $\Omega_i t =$(a)25, 35 and 40, respectively.

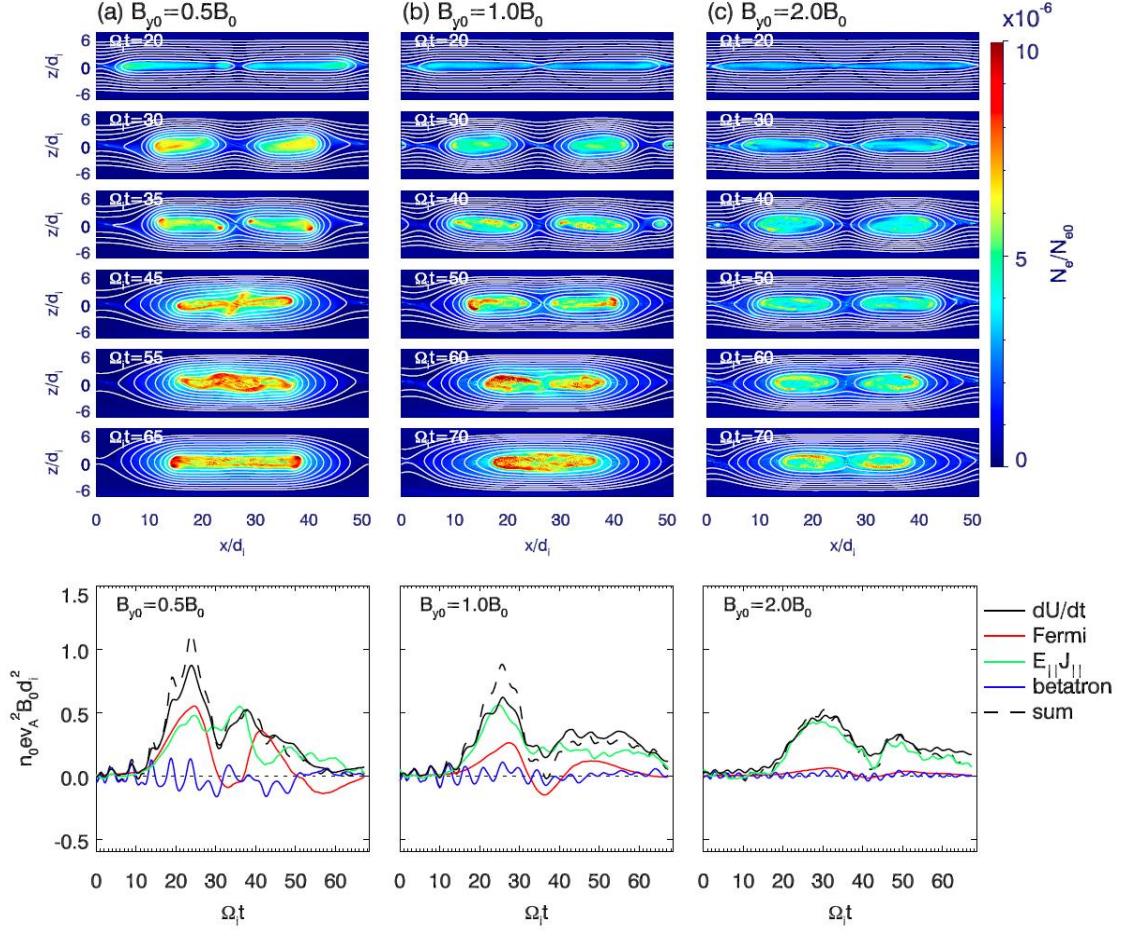

Figure 4. The evolution of the spatial distribution of electrons with energy larger than $0.1 m_e c^2$ and the contributions of the parallel electric field, Fermi and betatron mechanisms to the enhancement of electron energy in the whole simulation domain with different initial guide fields (a) $B_{y0} = 0.5 B_0$, (b) $B_{y0} = 1.0 B_0$ and (c) $B_{y0} = 2.0 B_0$, respectively. The contributions of different mechanisms to electron acceleration are integrated over the simulation domain. In the figure, $N_e$ is the electron number with energy larger than $0.1 m_e c^2$ at each grid point, while $N_{e0}$ is the total number of electrons over the simulation domain.

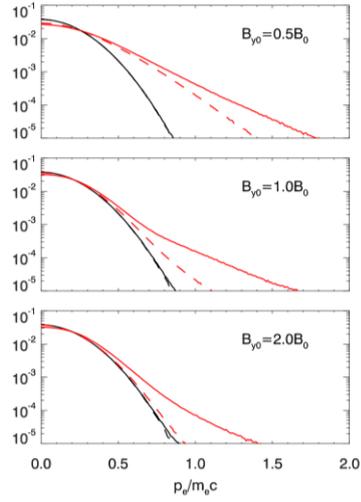

Figure 5. Parallel and perpendicular electron momentum spectra (over the entire domain) for simulations with guide field of $0.5B_0, 1.0B_0$ and $2.0B_0$, respectively. Solid lines correspond to the parallel momentum and the dashed lines represent the perpendicular momentum. The black and red lines represent $\Omega_i t = 0$ and $52$, respectively.